\documentclass[useAMS,usenatbib,usegraphicx]{journal2col}
\usepackage{graphicx}
\usepackage{amssymb}                                            
\usepackage{amsmath}                                            

\usepackage{natbib}

\def\aj{AJ}%
\def\araa{ARA\&A}%
\def\apj{ApJ}%
\def\apjl{ApJ}%
\def\aap{A\&A}%
\def\aaps{A\&AS}%
\def\azh{AZh}%
\def\mnras{MNRAS}%
\def\pasp{PASP}%
\def\pasj{PASJ}%
\def\nat{Nature}%
\def\iaucirc{IAU~Circ.}%
\newcommand{\ls}{\left(}
\newcommand{\ps}{\right)}

\newcommand{\rout}{r_{\mathrm{out}}}

\newcommand{\rg}{r_{\mathrm{g}}}
\newcommand{\rin}{r_{\mathrm{in}}}
\newcommand{\EBV}{E(\mbox{B}-\mbox{V})}
\newcommand{\ar}{a_\mathrm{r}}
\newcommand{\br}{b_\mathrm{r}}
\newcommand{\NH}{N_{\mathrm{HI}}}
\newcommand{\Mon}{\mbox{A\,0620--00}}
\newcommand{\GUMus}{\mbox{GS\,1124--683}}
\newcommand{\Md}{M_{\mathrm{d}}}
\newcommand{\Min}{\dot{M}_{\mathrm{in}}}
\newcommand{\hout}{h_{\mathrm{o}}}

\newcommand{\Section}[1]{\section{{\normalsize\textsc{#1}}}}

\title[Outburst Models
for Nova Monocerotis 1975 and Nova Muscae 1991]{Non-Steady State Accretion Disks in X-Ray Novae: Outburst Models
for Nova Monocerotis 1975 and Nova Muscae 1991}
\author[G. V. Lipunova and N.I. Shakura]
{G. V. Lipunova$^{1}$\thanks{E-mail: galja@sai.msu.ru}, 
N.I.Shakura$^{1,2}$\\
$^{1}$Sternberg Astronomical Institute, Universitetskiy pr., 13, Moscow 
119992, Russia\\
$^{2}$Max Planck Institut for Astrophysics, Garching, Germany
}
\begin{document}
\date{Received June 25, 2001; in final   form September 13, 2001}

\pagerange{\pageref{firstpage}--\pageref{lastpage}} \pubyear{2002}
\maketitle

\label{firstpage}

\begin{abstract}
 We fit outbursts of two X-ray novae (Nova Monocerotis 1975 = \Mon{} and
Nova Muscae 1991 = \GUMus{}) using a time-dependent accretion disk
model. The model is based on a new solution for a diffusion-type
equation for the non-steady-state accretion and describes the evolution
of a viscous $\alpha$--disk in a binary system after the peak of an
outburst, when matter in the disk is totally ionized. The accretion
rate in the disk decreases according to a power law. We derive formulas
for the accretion rate and effective temperature of the disk. The model
has three free input parameters: the mass of the central object $M$, the
turbulence parameter $\alpha$, and the normalization parameter $\delta
t$. Results of the modeling are compared with the observed X-ray and
optical B and V light curves. The resulting estimates for the turbulence
parameter $\alpha$ are similar: 0.2--0.4 for \Mon{} and 0.45--0.65 for
\GUMus, suggesting a similar nature for the viscosity in the accretion
disks around the compact objects in these sources. We also derive the
distances to these systems as functions of the masses of their compact
objects. \\ {\bf DOI}:
10.1134/1.1479424\newline~\newline~\newline~\newline
\end{abstract}

%
%
%

\section{Introduction}

 Accretion provides an efficient mechanism for energy release in stellar
systems, making many astrophysical objects observable. If the matter
captured by the gravitation of the central body possesses nonzero
angular momentum relative to this body, accretion  occurs in a disk. This
is true, for example, in binaries, where the angular momentum is
associated with the orbital rotation of the components. In the course of
accretion onto a compact object whose radius is comparable to the
gravitational radius, a substantial fraction of the total energy of the
accreted matter $mc^2$ is released. 

Outbursts reflect one of the most fundamental properties of accretion:
its non-steady-state character. Currently, a number of different models
are proposed to explain non-steady-state processes in accretion disks
and to describe the observed source variability. One problem is to find
an adequate description for the viscosity in the accretion disk:
viscosity is essential for the accretion, and the viscosity
characteristics specify the features of time-dependent
disk behavior.

In~\cite{lip-sha2000}, a new solution for the basic equation of
time-dependent disk accretion is found and applied to a model of an 
accretion  $\alpha$--disk around a compact star in a close binary. An
important
property of the disk in a binary system is that its outer radius is limited.
Angular momentum is carried away from the outer boundary
of the disk due to tidal forces, so that the rotation of outer parts of
the disk is synchronized with the rotation of the secondary. It is
assumed that the size of the accretion disk specified by the tidal
interactions is constant over the time interval considered. Another
assumption is that the rate of mass transfer from the secondary to the
disk is small compared to the accretion rate within the disk.

 The last condition is satisfied, for example, in an X-ray nova outburst.
The accretion rate in the disk during the outburst reaches tenths of the
Eddington rate $10^{-9} ~ (\vartheta\,M/M_\odot)~M_{\odot}/$yr or more,  where
$\theta$ is the accretion efficiency and $M$ is the mass of the central
object), while the observed rate of mass transfer from the
companion in quiescent periods is
$10^{-11}$--$10^{-12}~M_\odot/$yr~\citep{tan-shi1996, cherep2000}.  X-ray
novae are low-mass binaries containing a black hole or a neutron star
(see, for example,~\cite{cherep2000}). The other component, a low-mass
dwarf, fills its Roche lobe, so matter continuously flows into 
the disk~\citep{cherep2000}. 

Currently, more than 30 X-ray novae are known~\citep{cherep2000}.  Most
of them have light curves with similar exponentially decreasing
profiles~\citep{chen_et1997}. During the burst rise, 
the intensity increases by a factor  of
$10^2$--$10^6$ over several days, whereas the exponential decrease of
the light curve lasts for several months, with a characteristic time of
about 30--40 days.

Two  mechanisms for X-ray nova outbursts are developed: disk
instability and unstable mass transfer from the secondary. A final
choice between them has not been made, and each model faces some
problems (see, for example, \cite{cherep2000}). In 
disk-instability models, during the
outburst, the central object accretes matter accumulated by the disk
over decades of the quiescent state. This idea is supported by the fact
that the mass-transfer rate in quiescent periods is comparable to an
accretion rate corresponding to the 
outburst energy divided by the time between outbursts~\citep{tan-shi1996}. 

In any case, an important point here is that we assume the presence of a
standard disk at the time of maximum brightness of the source, as
confirmed by spectral observations~\citep{tan-shi1996}. By "standard
disk" we mean a multi-color $\alpha$--disk whose inner radius coincides
with the last stable orbit around the black hole, with the velocity of
radial motion of gas being small compared to other characteristic
velocities in the disk. Then, the solution of~\cite{lip-sha2000} can be
applied to the decay of an X-ray nova outburst. The light curve
calculated using the solution describes observations when the
contribution of the accretion disk dominates in the soft X-ray radiation
of the system. This phase is characterized by a definite spectral state
and can be distinguished in the evolution of an X-ray nova. 

Modeling the light curves of an X-ray nova in several X-ray and optical
bands enables one to derive the basic parameter of the disk, the
turbulence parameter $\alpha$ (this is a new, independent method for
determining $\alpha$ in astrophysical disks), as well as the
relationship between the distance and mass of the compact component.

Here, we apply our model to Nova Monocerotis \Mon{}, which
is the brightest nova in X-rays observed to the present time, and Nova
Muscae \GUMus.
\footnote{Observations collected from the literature and presented 
in uniform units can be found  at http://xray.sai.msu.ru/\~~\!galja/xnov/}

Currently, the most likely mechanism for turbulence and angular-momentum
transfer in accretion disks is thought to be Velikhov--Chandrasekhar
magnetic--rotational  instability~\citep{velikhov1959, chandra1961},
which is investigated in application to accretion disks 
by~\cite{bal-haw1991}. Calculations suggest that this type of instability
corresponds to $\alpha\sim10^{-2}$. 

The parameter $\alpha$, which was introduced by \cite{shakura1972},
describes large-scale turbulent motions.  The large-scale development of
MHD turbulence has been simulated, for example, by~\cite{armita1998} and
\cite{hawley2000}, and it was estimated  that $\alpha \sim 10^{-1}$. 

On the other hand, the following estimates for $\alpha$ were derived by
comparing theory and observations: for dwarf novae \hbox{$\sim 0.1$} during 
an outburst and \hbox{$\sim 0.02$} in a quiescent state 
in a model with a limit-cycle instability  (see, for
example,~\cite{canniz1988});  \hbox{$\sim 10^{-2}$} for disks in
galactic nuclei~\citep{sie-cze1989}; \hbox{$\sim 1$} for  Sgr~A$^*$ in
an advection-dominated model~\citep{naraya1995a}; and \hbox{$\sim
0.1-0.3$} in the inner,  hot advective part of the disk for the X-ray
novae \GUMus, \Mon, and V404~Cyg, basing on spectra in the low
state~\citep{naraya1996}.

\Section{MODEL FOR ACCRETION DISKS IN X-RAY NOVAE\label{xnov-model}}
The evolution of a viscous accretion disk is described by the
diffusion-type nonlinear differential equation~\citep{filipov1984}: 
\begin{equation}
\frac{\partial F}{\partial t}=
D\,\frac{F^m}{h^n}\,\frac{\partial^2F}{\partial h^2}\, ,
\label{nonlin}
\end{equation}
where $F=W_{r\varphi}\, r^2$ is the total moment of the viscous forces acting
between adjacent rings of the disk divided by $2\pi$, $W_{r\varphi}$ is the 
component
$w_{r\varphi}$ of the viscous stress tensor integrated over the thickness of the
disk, $h = \sqrt{GMr}$ is the specific angular momentum, and $M$ is the mass
of the central object. The dimensionless constants $m$ and $n$ depend on the
type of opacity in the disk. If the opacity is determined largely by
absorption (free--free and bound--free transitions), then $m=3/10$ and
$n=4/5$. The ``diffusion coefficient" $D$ specified by the vertical
structure of the disk relates the surface density $\Sigma$, $F$ , and
$h$~\citep{filipov1984, lyub-shak1987}: 
\begin{equation}
\Sigma = \frac{(GM)^2\, F^{1-m}}{2\,(1-m)\,D\, h^{3-n} }\, .
\label{SigDF1}
\end{equation}
Relation~(\ref{SigDF1}) is derived from an analysis of the vertical structure of
the disk. 

A class of solutions for Eq.~(\ref{nonlin}) is derived
in~\cite{lyub-shak1987} during studies of the evolution of a torus of
matter around the gravitating center under the action of viscous forces,
which are parameterized by the turbulent viscosity parameter $\alpha$
introduced in~\cite{shakura1972}. In particular, a solution was obtained
for the stage when the torus has evolved to an accretion-disk
configuration, from which matter flows onto the central object. When the
accretion rate through the inner edge decreases, the outer radius of the
disk simultaneously increases---the matter carries angular momentum away
from the center. In this model, the accretion rate decreases with time
as a power law. The power-law index depends on the type of opacity in
the disk. 

An important property of a disk in a binary system is the cutoff of the
disk at the outer radius, in the region where the angular momentum is
carried away due to the orbital motion (see, for
example,~\cite{ich-osa1994}). Taking into account the corresponding
boundary conditions, we obtained a new solution for Eq.~(\ref{nonlin})
in a general form for a disk with uniform opacity~\citep{lip-sha2000}.
We described the vertical structure using calculations
of~\cite{ket-sha1998} in the framework of the generally accepted
$\alpha$-disk model~\citep{sha-sun1973}. Following~\cite{lyub-shak1987},
we considered two opacity regimes: with the dominant contribution to the
opacity made by the Thomson scattering of photons on free electrons, and
by the free--free and bound--free transitions in the plasma. As a
result, we obtained explicit expressions describing the time variations
of the physical parameters of the accretion disk. The solution describes
the evolution of the accretion disk in a binary during the decay of the
outburst, while the matter in the disk remains completely ionized. The
accretion rate decreases with time according to a power law; however, in
this case, the power-law index is larger than that for the solution
of~\cite{lyub-shak1987}: $-5/2$ compared to $-19/16$ when Thomson
scattering dominates and $-10/3$ compared to $-5/4$ when absorption
dominates. The solutions in the two opacity regimes join smoothly,
providing a basis for applying a combined solution to describe the
evolution of a disk with a realistic opacity. Our study of disks in
stellar binary systems indicates that the second opacity regime is
realized during times investigated\footnote[4]{{\em Note added to the
astro-ph version}: we point out that important here is the type 
of opacity in the outer disk, where the bulk of the mass is contained.}. 

The law for the variation of the accretion rate can be easily derived
from the condition of mass conservation in the disk. Let a Keplerian
$\alpha$--disk in a binary have fixed inner and outer radii $\rin$ and
$\rout$. The mass of the disk is $\Md =\int\limits_{\rin}^{\rout}
2\,\pi\, r\, \Sigma \, \mathrm{d}r$. If this mass varies only due to
accretion through the inner disk boundary (i.e., the accretion rate onto
the outer boundary is substantially lower), then $\mathrm{d}\Md /
\mathrm{d} t = \Min (t)$.  Let the disk parameters $F$ and $\Sigma$ be
represented as the products $F(t)\,f(\xi)$ and
$\Sigma(t)\,\sigma(\xi)$,  where $f(\xi)$ and $\sigma(\xi)$ are
dimensionless functions of the radial coordinate $\xi=h/\hout$. The
value of $\hout$ is fixed and equal to the angular momentum at the outer
disk boundary. Let us express the surface density $\Sigma$ in the
integrand in terms of $F$ using Eq.~(\ref{SigDF1}). It follows from the
equation for  the angular momentum transfer that  $\Min (t)=
-\,2\,\pi\,F(t)\,f'(\xi)/\hout$~\citep{lip-sha2000}; therefore we obtain
\begin{multline}
\Md =  - \, \Min^{1-m} \,\int\limits_{\rin}^{\rout}\,
\ls\frac{\hout\,f(\xi)}{2\,\pi\,y(\xi)}\ps^{1-m} \frac{\pi\,(G\,M)^2}
{(1-m)\,D\,h^{3-n}}\, r\,\mathrm{d}r\,  \\=\,
 (\mathrm{d}\Md / \mathrm{d} t )^{1-m}\times const\, ,
\label{Mdisk}
\end{multline}
where dimensionless $y(\xi)=f'(\xi)$. We will integrate Eq.~(\ref{Mdisk})
over $t$. Using the fact that $m=10/3$ for the Kramers law, we conclude
that \hbox{$\Md\propto (t+\delta\,t)^{-7/3}$} and \hbox{$\Min \propto
(t+\delta\,t)^{-10/3}$}. Note that, to produce such time dependence, it is
sufficient that the ``diffusion coefficient" $D$ is constant in time,
being a function of the radius. 

The bolometric luminosity $L=\theta \Min c^2$ varies according to the
same law as the accretion rate through the inner disk radius. As shown
in~\cite{lip-sha2000}, the exponential light curve of an X-ray nova can
be then explained if the considered spectral band contains the flux
integrated over the exponential falloff in the Wien section of the disk
spectrum rather than a fixed fraction of the bolometric luminosity. 

In the model considered, the accretion rate depends on the distance from
the center. In the central regions,  the accretion rate is
virtually independent of the radius; but this is not true for the outer
disk. The bolometric flux from regions of
the disk between rings with radii $0.1\,\rout$ and $\rout$ is about
$6$\% lower than the values obtained for a stationary standard disk
model. The optical flux from the time-dependent disk differs from that
from a stationary disk by a smaller amount, depending on a size of the 
region producing the optical flux. This size is specified by the mass of
the central object and the accretion rate in the disk.

 To take into account the effects of general relativity in the vicinity
of the compact object, we use a modified Newtonian gravitational
potential in the form suggested by~\cite{pacz-wiita1980}. For a
Schwarzschild black hole, this potential is 
\begin{equation}
\psi = \frac{G\,M}{r-\rg}~,
\label{Pacz}
\end{equation}
where $\rg = 2\,G\,M/c^2$. The
accretion efficiency $\theta$ for this potential is a factor of $\approx 1.45$
smaller than that for a Newtonian potential. 

\Section{MODELING PROCEDURE\label{sect_proc_model}}

\subsection{Derivation of Theoretical Curves}

 In the model for a time-dependent disk in a binary
system~\citep{lip-sha2000}, with the opacity in the outer disk defined
by the bremsstrahlung absorption, the variation of the accretion rate in
the disk as a function of time is given by the formula 
\begin{equation}
\dot M(h,t) = -2\,\pi\, \frac{1.224\,y(h/h_\mathrm{o})}{h_{\mathrm{o}}}\,  \left( \frac{h_\mathrm{o}^{14/5}}
           {D\, (t+\delta t)} \right) ^ {10/3}\,       ,
\label{rate_ff}
\end{equation}
where $t$ is the time, $\delta t$, a normalizing shift, $h$, the specific angular
momentum, $h_{\mathrm{o}}$, the specific angular momentum of the matter at the 
outer
radius of the disk, $D$, the constant in Eq.~(\ref{nonlin}), and 
\begin{equation}
y(\xi) \approx 1.43 - 1.61\,\xi^{2.5}+0.18\, \xi^5\, .
\end{equation}

\begin{table}
\caption[Input model parameters.]{Input model parameters. 
For parameters denoted with "+" we use observed values.}
\begin{center}
\begin{tabular}{c|l|c}
\hline
\multicolumn{1}{c|}{(1)}&\multicolumn{1}{c|}{(2)}&\multicolumn{1}{c}{(3)}\\
\hline
$M$& Mass of the central object&\\
$M_{\mathrm{o}}$ & Mass of the optical component & $+$\\
$P$& Orbital period of the binary&$+$\\
$f(M_\mathrm{o})$&Mass function of the optical component& $+$\\
$\alpha$ &Turbulence parameter in the disk&\\
$\NH$ &Number of H atoms per cm$^2$ to the source& $+$\\
$\mu$ & Molecular weight of gas in the disk $0.5$&\\
$\delta t$ & Inner parameter of the model&\\
\hline
\end{tabular}
\end{center}
\label{input_parameters}
\end{table}

Table~\ref{input_parameters} presents the input parameters for the 
model. Three parameters are free: the mass of the compact object $M$,
the turbulence parameter $\alpha$, and the normalizing parameter $\delta t$. A
change of the mass of the optical component $M_{\mathrm{o}}$ affects the
disk size, which affects the rate of variation of the accretion rate in
the disk. The following values can be obtained from the input
parameters: 

(1) The system semiaxis $a$, calculated  as 
\begin{equation}
 a = \ls\frac{G\,(M+M_{\mathrm{o}})\,P^2}{4\,\pi^2}\ps^{1/3}\, ,
\label{half-axe}
\end{equation}
assuming that the orbits are circular.

(2) The system inclination 
\begin{equation}
i = \arcsin { \ls\left[ \frac{f(M_\mathrm{o})\,(1+q)^2}{M\,q^2} \right]^{1/3}\ps
}\, ,
\label{inclination}
\end{equation}
where the mass ratio is $q=M/M_{\mathrm{o}}$.

(3) The radius of the last nonintersecting orbit around the primary,
which depends on the mass ratio of the binary components and, in
general, does not exceed 0.6 of the Roche lobe size (values are
tabulated in~\cite{paczynski1977}); this corresponds to the radius of the
outer boundary of the disk $\rout$~\citep{paczynski1977, ich-osa1994}.

 (4) The
diffusion coefficient $D$ appeared in Eqs.~(\ref{nonlin}) and (\ref{rate_ff}): 
\begin{multline}
D = 5.04\times 10^{34}\,\alpha^{4/5} \,(\mu/0.5)^{-3/4} \, (M/M_\odot) \,
B_\mathrm{f}~~\\ [\mbox{g}^{-3/10}\, \mbox{cm}^{5}\,\mbox{s}^{-16/5}] \, ,
\label{dif_const_ff}
\end{multline}
where $B_\mathrm{f}=(\Pi_1^{1/2}\,\Pi_2\,\Pi_3^8\,\Pi_4)^{-1/10}$ is a
combination of dimensionless parameters specified by the vertical
structure of the disk, calculated and tabulated as functions of the
optial depth in~\cite{ket-sha1998}\footnote{In~\cite{ket-sha1998}, 
Table 1b should read $\Pi_4$ instead of $\Pi_3$ ; the 5th column should
be ignored. In~\cite{lip-sha2000}, there is a misprint in $D$ in (26)
and (31).}. Thus, $B_\mathrm{f}$ depends on the variable optical depth, 
which can be obtained from the disk characterstics at the radius.
We found that $B_\mathrm{f}$ depends stronger on time than on radius, 
and the time dependence is also not very strong, since $B_\mathrm{f}$ is the combination of
the parameters raised to powers much smaller than unity. In the modeling, we
adopt $B_\mathrm{f}\,$ calculated iteratively at the half-radius
of the disk and at the middle of the time interval investigated.

 To calculate the effective temperature in the disk, we use the formula
\begin{multline}
T^4(r)= -\frac{\dot M}{4\, \pi\, \sigma}\, \omega(r)\, r\,
\frac{\mathrm{d} \omega(r)}{\mathrm{d}r} \,
\ls1-\frac{\omega(r_\mathrm{in})}{\omega(r)}\,
\ls\frac{r_\mathrm{in}}{r}\ps^2\ps \times \\
\times\sqrt{\frac{\rout}{r}}\,\frac{f(\sqrt{r/\rout}) }{f'(0)}\, ,
\label{T(r)_combined} \end{multline} where $f(\xi)=1.43\,\xi
-0.46\,\xi^{7/2}+0.03\,\xi^6$, $\omega(r)$ is the angular velocity in
the disk (which is Keplerian away from the compact object), $\sigma$ is
the Stefan--Boltzmann constant, and $\rout$ is the radius of the outer
boundary of the disk. In Eq.~(\ref{T(r)_combined}), $\dot M$ is equal to
the value $\dot{M}(0,t)$ defined by Eq.~(\ref{rate_ff}). The central
regions of the disk (where $r\ll\rout$ and the product of the last two
factors in Eq.~(\ref{T(r)_combined}) yields approximately unity) produce
the largest contribution to the X-ray emission; here, the accretion rate
is nearly constant over radius, and the distribution of the effective
temperature essentially coincides with that in a stationary disk. We
also take into account non-Newtonian nature of the metric around the
compact object in the central regions of the disk. For the
Paczynski--Wiita potential (\ref{Pacz}), one has 
\begin{equation}
\omega(r) = \sqrt{\frac{G\, M}{r}}\,\frac{1}{r - \rg}\, .
\label{angular_Pacz}
\end{equation}
We assume that the bulk of the optical flux comes from the disk (at
distances \hbox{$r\gg r_{\mathrm{in}}$}), while the radiation from the
``transition layer'' at the outer boundary, where the momentum is
carried away, is significantly less.

 The spectral flux density detected by an
observer can be calculated using the following formula~\citep{bochkarev_et1988e}
\begin{equation}
F_\nu = \frac{L_\nu\, \cos{i}}{2\,\pi\,d^2} \, \exp{(-\tau_\nu)}\,
~[\mbox{erg/cm$^2$\,Hz\,s}]\,,
\label{spec_plot_pot}
\end{equation}
where $d$ is the distance to the system, not set {\it a-priori} in the
model, $\tau_\nu$ is the optical depth corresponding to the 
absorption toward the source, and $L_\nu$ is the spectral luminosity of
two sides of the disk according to the formula 
\begin{equation}
L_\nu =2\,\pi\,\int\limits_{r_\mathrm{in}}^{\rout}
{B(\nu,T(r)) \,2\,\pi\,r\,\mathrm{d}r }~
[\mbox{erg}/\mbox{c}\,\mbox{Hz}]\, ,
\label{fluxes}
\end{equation}
where $B(\nu,T(r))$ is the Planck function.

The model curves $F_i$ (erg/cm$^2$s) are calculated by integrating
$F_\nu$  over frequency. At X-ray energies, $\tau_\nu =  \sigma(\nu) \,
\NH \,$, and the cross section for  absorption by the hydrogen atoms
$\sigma(\nu)$ is presented in the form of an  analytical
spline~\citep{mor-mcc1983}. The number of hydrogen atoms in the line of sight
toward the source $\NH$ can be found in the literature, and can also be
calculated using the approximating formula~\citep{zombeck1990} 
\begin{equation}
\NH \approx 4.8\times 10^{21}\, \EBV\,~\mbox{\rm{atoms}}/
\mbox{\rm{cm}}^{2} \mbox{\rm{mag}}\, ,
\label{hydrogen}
\end{equation} 
provided the color excess $\EBV$ is known and assuming that the
main contribution to the absorption is made by the hydrogen atoms of the
interstellar medium and not by those directly related to the source. 

For the optical spectrum, absorption is taken into account after
integrating $F_\nu$ over frequency. We use the interstellar absorption 
$A_\lambda = 2.5 \, \tau/\ln 10$ instead of the optical depth 
$\tau_\nu$; here $\tau$ is some effective value for the optical band denoted
by $\lambda$.

We calculate the light curves $F_i$ in a chosen time interval 
$t\in[t_1,t_2]$, and we set $t=0$ at the peak of the outburst. The accretion
rate at the inner radius, bolometric luminosity of the disk, and fluxes
in specified spectral intervals (the X-ray and the optical B and V
bands) are calculated with a step of \hbox{$\Delta t=1$~day}.

\subsection{Comparison of the Model and Observed Light Curves}

We use the light
curves in one X-ray band and in the B and V band.

Spectral observations of X-ray novae can help to determine in which
timporal and spectral intervals the flux is dominated by the disk
radiation, with much less contributions from other components.
\label{x-ray-choice}In general, it is obviously desirable to use data
at the softest X-ray energies (below $\sim 10$~keV), since the typical
X-ray spectrum of an X-ray nova after the outburst peak is a combination
of the disk spectrum (modeled as emission from a multi-color black-body
disk) and a harder power-law component~\citep{tan-shi1996}. For each
source, we must consider the spectral evolution and exclude time
intervals when either the contribution from the nondisk components (or
power-law components) is substantial or the disk spectrum differs
appreciably from that for a multi-color disk model. Nondisk
components in the emission can alter the slope of the decaying light
curve, and this slope dramatically affects the resulting values of
$\alpha$. 

We reduce observed stellar magnitudes at optical wavelengths to
fluxes in erg/(cm$^2$s) using the formula 
\begin{equation}
 F_i=\Delta \lambda_i\, 10^{-0.4\,m_i-A^0_i}\, ,
 \label{magn_to_erg}
\end{equation}
where the zero-points $A^0_i$ and effective bandwidths $\Delta \lambda_i$ are
presented in Table~\ref{opt_consts}.
\begin{table}
\caption{Zero-points, central wavelengths, and effective widths for the
optical bands~\citep{zombeck1990}.
The corrected value is denoted by $^*$; 
the value from~\citet{zombeck1990} is given in paren╜theses 
(see explanation in the text after Eq.~(\ref{magn_to_erg})).} 
\begin{center}
\begin{tabular}{c|c|c|c}
\hline
 \multicolumn{1}{c|}{ Band}
& \multicolumn{1}{c|}{ $A^0_i$}
& \multicolumn{1}{c|}{$\lambda$, \AA}
&\multicolumn{1}{c}{$\Delta \lambda_i$, \AA }\\
\hline
U&$8.37$&	$3650$&$680$\\
B & $8.198^*$ ($8.18$)	&$4400$& $980$\\
V&  $8.44$ &$5500$&$890$\\
\hline
\end{tabular}
\label{opt_consts}
\end{center}
\end{table}
 We have corrected $A^0_\mathrm{B}$ values in accordance with
\cite{bochkarev_et1991}, where the disk color indices are calculated
using the radial temperature distribution of a standard disk, taking
into account the actual transmission functions for the optical filters.
The values calculated with our code, using rectangular band-pass for the
optical filters, coincide with those from ~\cite{bochkarev_et1991} to
within 0.01$^m$. To achieve this agreement, it has been necessary to
correct the zero-point in the B band. 

Let us consider the disk flux $F_\mathrm{x}^\mathrm{obs}$ observed in
some X-ray spectral band at some time $t\in[t_1,t_2]$. We can derive the
distance $d$ to the source for the chosen model parameters (see
Table~\ref{input_parameters}) by comparing observed flux with
theoretical $F_i(t)$, taking into account the absorption toward the
source and the inclination $i$ calculated with~(\ref{inclination}). We
assume that the disk is coplanar to the binary orbit. 

Further, we calculate model light curves $F_i$ using the derived $d$. We
find the color excess $\EBV^\mathrm{model}$ that is required to obtain
agreement with the observed optical flux in the B band. Each optical
light curve is then corrected for this $\EBV^\mathrm{model}$ by
substructing value $A_\lambda/2.5$ from the logarithm of the optical
flux. For this, we use the values $R_\mathrm{B}=4.2$,
$R_\mathrm{V}=3.2$, $R_\mathrm{U}=5$~\citep{zombeck1990} in the formula
relating the color excess and the interstellar absorption: 
\begin{equation}
A_\lambda=R_\lambda\,E(\mathrm{B}-\mathrm{V})\, .
\label{reddening}
\end{equation}

Observations show that certain portions of X-ray light curves of X-ray
novae are very close to exponential dependences and, thus, can be
described by only two parameters. 
We approximate the observed light curves with straight lines $\log F_i =
\ar\,t\, +\,\br$, whose slopes $\ar$ are to be compared to those of the
theoretical curves at some point $t \in [t_1,t_2]$ in the middle of the
time interval. For three spectral bands, we fit fluxes at this
$t$, taking into account the observed color excess $\EBV$ and the color
of the disk B--V. A set of models is selected to provide a certain
agreement between the X-ray and optical fluxes and slopes of the
observed and theoretical light curves. The value of $\chi^2$ is
calculated for the selected models. 

A large scatter of the observed disk color B--V, which reaches 0.05
magnitudes (see below), can be caused by both observational errors and
possible short-term fluctuations of the disk color (due to precession,
the superposition of the optical flux from the secondary, etc.) that we
do not examine here. We also do not take into consideration the
contribution of radiation from the central X-ray source reradiated from
the outer parts of the disk. 

For some of the light curves, the overall traces are rather close to
exponential, but the exponential law does not agree with the
observations by $\chi^2$ criterion (at the 0.05 significance level),
possibly due to an underestimation of the observational errors, a
superposition of fluctuations in emission, and model limitations.

\Section{MODELING X-RAY
NOVA \hbox{A\,0620--00} IN THE 3--6 keV X-RAY AND B AND V OPTICAL BANDS}

\subsection{Observations}

\subsubsection{X-ray Light Curves}

 The nova outburst in
Monoceros in 1975 (\Mon, \hbox{V 616 Mon}) was observed in X-rays with
the orbiting observatories Ariel-5, SAS 3, Salut 4, and
Vela 5B~\citep{elvis_et1975, doxsey_et1976, kurt_et1976, kaluz_et1977,
tsunem1989}. \Mon{} was the first X-ray transient
identified with an optical burst~\citep{Matilsky_et1975,Boley_et1976}. 

We take the 3--6 keV data of~\cite{kaluz_et1977}, obtained with the
Ariel-5 All Sky Monitor.  Following~\cite{chen_et1997}, we assume that
the peak of the outburst was on Aug 13, 1975, or JD 2442638.3. The
corresponding data in Crab flux units are taken in the electronic form from
the HEASARC (High Energy Astrophysics Science Archival Research Center) 
database~\citep{chen_et1997}\footnote{ftp://legacy.gsfc.nasa.gov/FTP/heasarc/dbase/misc\_files/xray\_nova/
\label{ftp-address}}.

\begin{figure} 
  \begin{center}
   \resizebox{!}{0.4\textwidth}{\includegraphics{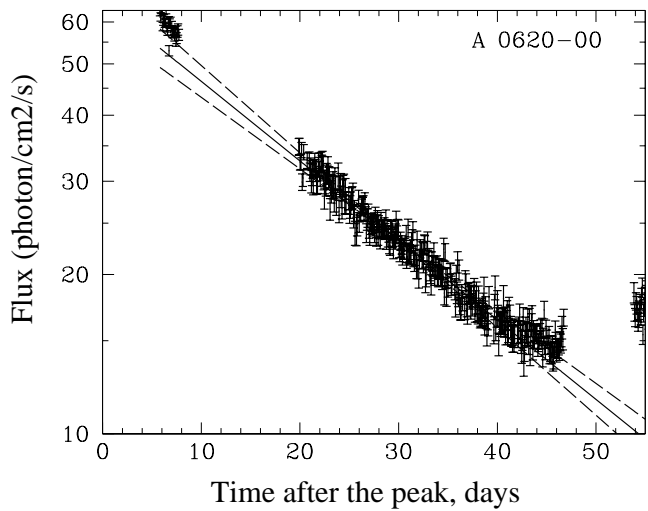}}
  \end{center}
  \caption{Straight-line approximation of the \Mon{}  X-ray 
light curve. The regression line for the data in the interval
$t\in[20,40]$~days (solid line) and selection 
boundaries for the theoretical curves (two dashed lines) are plotted. 
The observational data are from \citet{kaluz_et1977}.
}  \label{approximation0620X}
\end{figure}
We fit the X-ray light curve in units of photons/cm$^2$s.
Figure~\ref{approximation0620X} illustrates selection of models 
according to the slope of the 3--6~keV X-ray light curve at $t=30$~day.
The regression line constructed using observations in the interval
$t\in[20,40]$~day has $\ar=-0.01502\pm 0.0002$ and $\br=1.816\pm
0.007$. The data at $t\approx 10$~days cannot be ascribed with
confidence to the exponential section of the light curve caused by the
disk radiation. The dashed curves indicate the boundaries within which
we select the models. Within these boundaries, the slopes of the lines
vary in the range $(0.9-1.1)\,\ar$. Value of $\chi^2$, calculated for
the observational data and the lines with such slopes, divided by the
number of degrees of freedom, does not exceed 1.3.

\subsubsection{X-ray Spectrum at the Time of the Outburst\label{spectrum0620}}

 As has often been noted, X-ray novae in general and \Mon{} in
particular (see, for example,~\cite{kuulkers1998}) display softening
of the spectrum during the initial decay in the light curve. For \Mon, 
this was  pointed out, for example, by~\cite{carpenter_et1976} for
\hbox{3.0--7.6~keV} observations and \cite{citterio_et1976}, for
\hbox{3--9~keV}.

Spectral data with high resolution (10~eV at 2~keV and 285~eV at
6.7~keV) obtained with the Columbia spectrometer OSO~8 on board the
Ariel-5 spacecraft in October 1975 were analyzed
in~\cite{long-kest1978}. The X-ray continuum of \Mon{} on September
17--18, 1975 (the 34--35th day after the peak) was best fit by a
blackbody model with $kT\sim 0.5$~keV. 

Note that, at the energies of interest to us ($>1$~keV), the spectrum of
a multi-color, blackbody disk can be approximated by a Wien
spectrum:
\begin{equation}
I_E =  \frac{2}{c^2 \, h^3}\, {E^3}\,{\exp{(-E/kT_\mathrm{max})}}  \ ,
\label{Wien}
\end{equation}
where $T_\mathrm{max}$ is the maximum temperature in the disk, which
occurs at the radius $\approx1.58 \, \rin$  in the case of the
Paczynski--Wiita gravitational potential (\ref{Pacz}). In a Newtonian
potential approximation, the radius for the maximum temperature is
approximately $1.36 \, \rin$.

\paragraph*{Number of Hydrogen Atoms/cm$^2$ toward \Mon{} and the Color
Excess\label{hydr_section_0620}}
~\newline
 The number of hydrogen atoms was estimated from the falloff in the
soft X-ray spectrum due to the absorption of X-rays with energies
$<1$~keV~\citep{carpenter_et1976, doxsey_et1976, kurt_et1976}:
$3\times 10^{21}-10^{22}$~atoms/cm$^{2}$. 

In~\cite{wu_et1983}, the estimate $\EBV=0.35 $ was derived using UV
observations obtained with the ANS satellite. Twenty-five stars around
\Mon{} were studied, and a specific absorption curve for the region was
derived. Using (\ref{hydrogen}), one finds $\NH \sim 1.7\times 10^{21}$
atoms/cm$^{2}$. The total number of hydrogen atoms in the Galaxy in the
line of sight toward \Mon{} is $\sim4\times
10^{21}$~atoms/cm$^{2}$~\citep{weav-will1974}. 

Obviously, we cannot exclude the possibility of substantial and variable
absorption of the X-ray radiation in the source itself. However, the
character and origin of this variability are unknown. We model the
observations for Nova Monocerotis for a set of $\NH$ values in the range
$10^{21}$--$10^{22}$ atoms/cm$^{2}$. For the color excess, we adopt 
value $\EBV=0.35\pm0.01$~\citep{wu_et1983}.

\subsubsection{Optical Light Curves\label{0620_optic}}

Optical observations from~\cite{liutyi1976, shugarov1976, 
vandenbergh1976, duer-walt1976, rebertson_et1976, lloyd_et1977} are
used. We construct linear regression fits $\log F = \ar\,t\, +\,\br$ to
the B and V optical data in logarithmic flux units for times
$t\in[0,47]$ days using the weighted least-squares method. This yields
$\ar=-0.0079\pm 0.0002$, $\br=-9.675\pm 0.005\,$ for the B band and 
$\ar=-0.0071\pm 0.0002$, $\br=-9.885\pm 0.006\,$ for the V band (see
Section~\ref{x-ray-choice}). The reduced $\chi^2$ values for these fits
are roughly 12 and 43 for the B and V bands, with 102 and 89 degrees of
freedom, respectively. This suggests that the adopted errors for the
optical observations might be underestimated (for example, systematic
deviations have not been taken into account) and/or that our assumed
exponential (quasiexponential) decrease of the optical flux does not
yield a complete description of the observed curves, due to various
fluctuations and variations of the optical flux superimposed on the
overall trace of the light curve. Nevertheless, for modeling 
we assume that the basic
trend of the light curves can be fit by a quasi-exponential decay. 

In model fitting, we adopt the color of the disk at $t=30$~days
\hbox{B--V}$=0.24\pm 0.03^m$, derived from the above observational data.

\subsection{Results for \Mon\label{results0620}}
 We compare the theoretical and observed curves for $t\in[20,40]$~days
after the outburst peak. Table~\ref{input_parameters_A0620} summarizes
the parameters attempted. The number of hydrogen atoms along the
line of sight toward \Mon{} does not appreciably affect the results,
since the absorption at 3--6~keV is small.

\begin{figure} 
  \begin{center}
 \resizebox{!}{0.55\textwidth}{\includegraphics[trim= 3cm 6cm 2cm 5cm,clip]{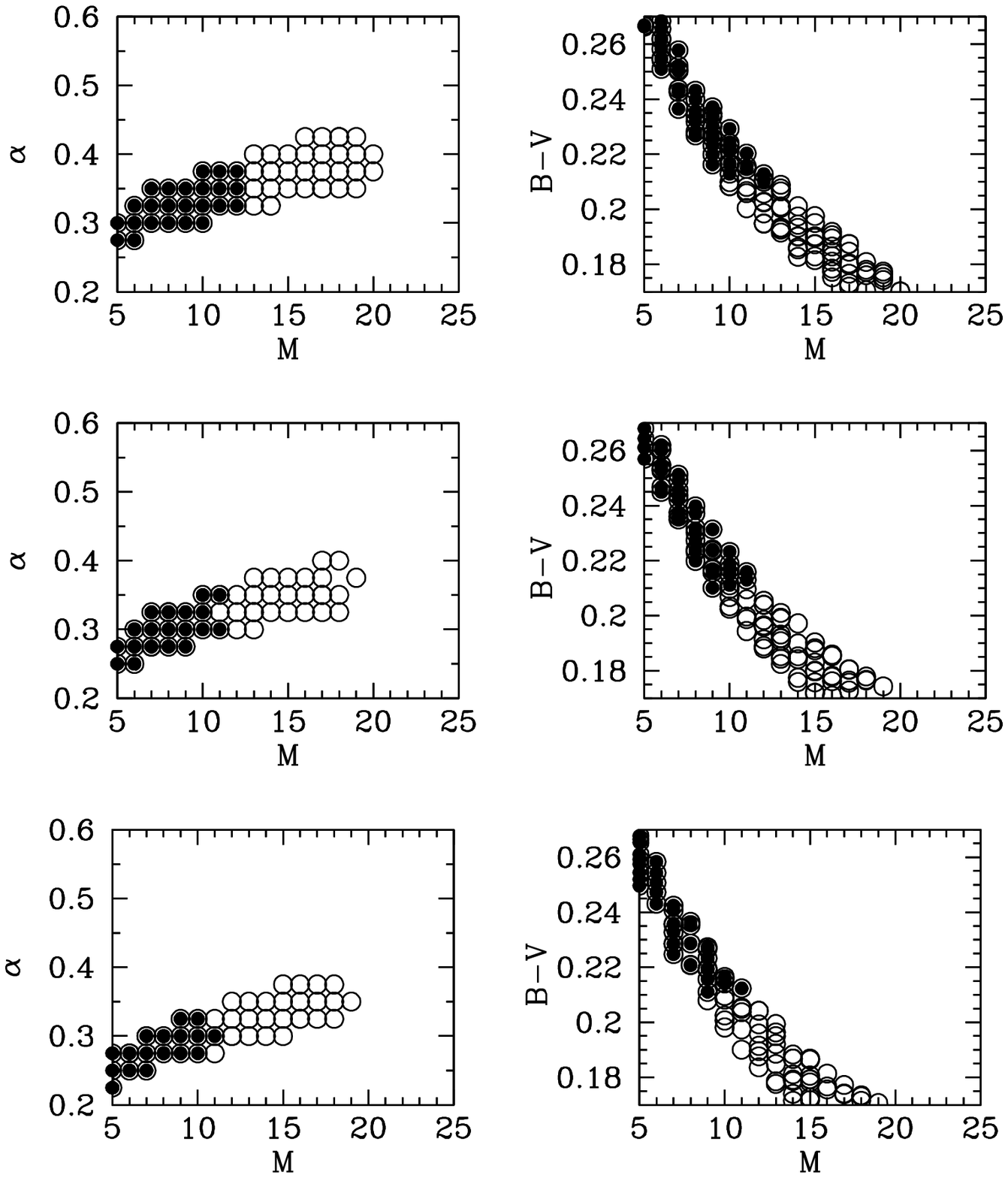}}
  \end{center}
  \caption{Modeled $\alpha$ and color of the \Mon{} disk. 
Results are given for the values from
Table 3. The mass of the optical component varies from top to bottom: 
0.3, 0.5, and 0.7~$M_\odot$. The filled circles satisfy the disk
color condition $B-V = 0.24 \pm╠ 0.03$~mag for $t = 30$~days.
}
  \label{models0620_3}
\end{figure}
Figure~\ref{models0620_3} presents the results of our modeling for the
parameters from Table~\ref{input_parameters_A0620}. We can see that
$\alpha$ lies in the range 0.225--0.375 (for slightly different masses
$M_\mathrm{o}$). In~\cite{lip-sha2001}, we adopted
$M_\mathrm{o}=0.5\,M_\odot$ and $\EBV=0.39$ and used a broader range of
slopes than in Fig.~\ref{approximation0620X}, so a broader interval for
$\alpha$ was obtained for \Mon{} in that study, from 0.3 to 0.5.

\begin{figure} 
  \begin{center}
   \resizebox{!}{0.35\textwidth}{\includegraphics{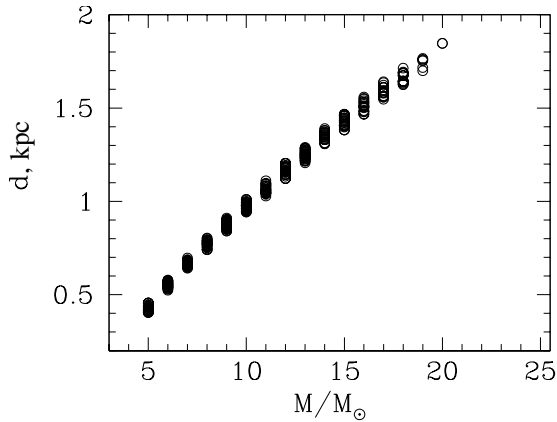}}
  \end{center}
  \caption{ The distance-black hole mass dependence for \Mon{} 
for the model parameters from Table 3.}
  \label{d-m0620}
\end{figure}
\begin{figure} 
  \begin{center}
   \resizebox{!}{0.35\textwidth}{\includegraphics{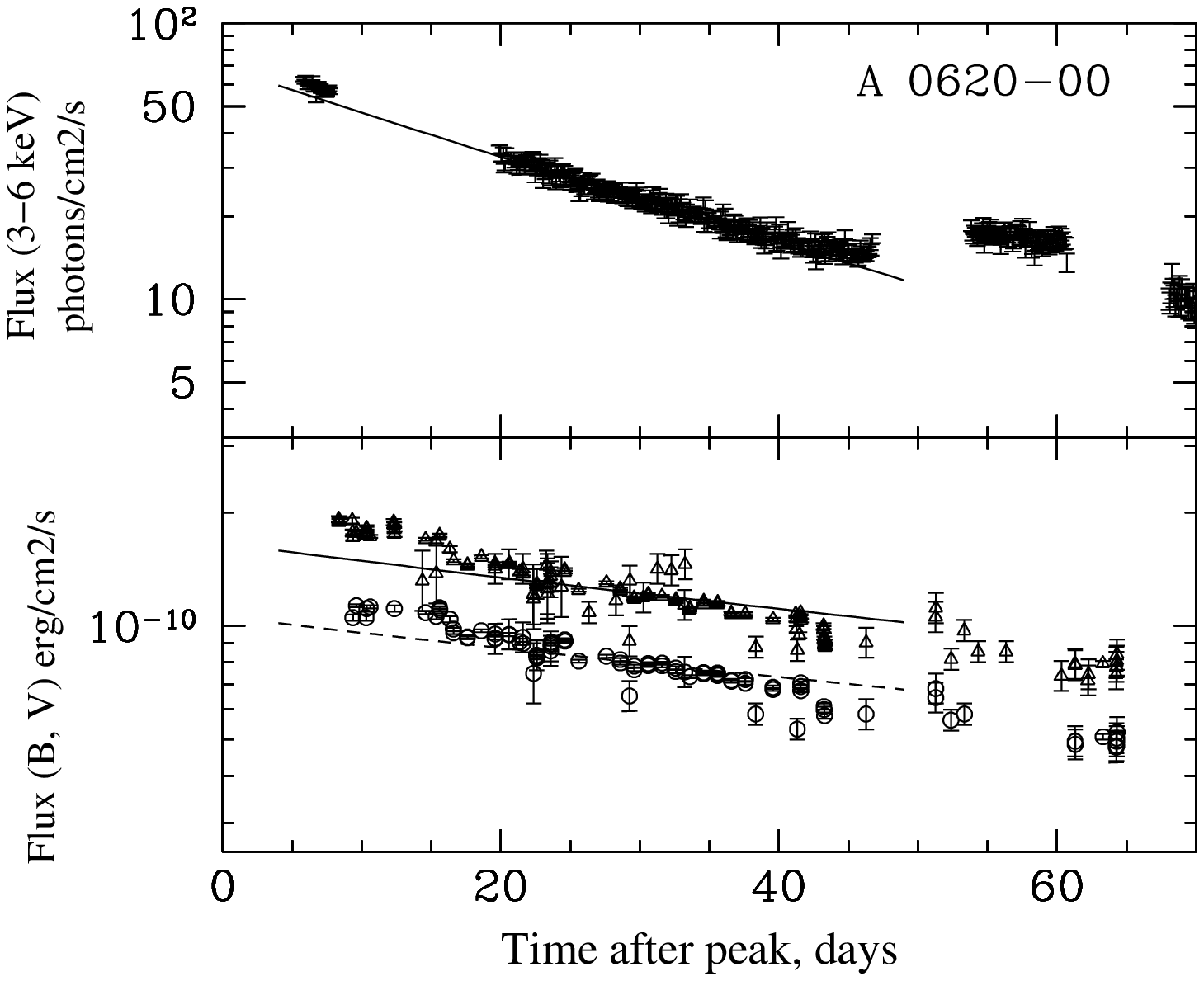}}
  \end{center}
  \caption{Example of modeling light curves of \Mon{} in the 3--6~keV, 
B, and V band. The model parameters are $M = 7M_\odot$,
$M_o = 0.5M_\odot$, $\alpha = 0.3$, $\delta t = 168$~days, 
and $N_\mathrm{HI} = 3 \times 10^{21}$~atoms/cm$^2$.
}
  \label{model0620-1}
\end{figure}

Figure~\ref{d-m0620} shows the resulting relationship
between the distance to \Mon{} and the mass of the black hole. The
distance to \Mon{} has been estimated to be from 0.5 to 1.2 kpc
(for example, in~\cite{oke1977, vandenbergh1976}; see also  
reviews by~\cite{chen_et1997, cherep2000}). Figure~\ref{d-m0620}
indicates that for $d\sim0.9$~kpc~\citep{oke1977} the mass of the black 
hole in \Mon{} is $\sim 9\, M_\odot$. This mass is consistent with
previous estimates~(see~\cite{cherep2000} and references
therein).
\begin{table}
\caption[Параметры моделей \Mon]{ 
Input parameters for \Mon{} models. 
The mass of the optical component $M_{\mathrm{o}}=0.5\, M_{\odot}$, 
its mass function, and
the binary period are taken from~\cite{chen_et1997, cherep2000}.
}
\begin{center}
\begin{tabular}{c|l}
\hline
Parameter  &  Tested values \\
\hline
$M$              & $5-25\, M_\odot$                   \\
$M_{\mathrm{o}}$ &  $0.3$, $0.5$, $0.7\, M_\odot$                \\
$P$              &  0.322 day                      \\
$f(M_\mathrm{o})$&  $2.7 \, M_\odot$         \\
$\alpha$         &  $0.1-1$                        \\
$\NH $ & $3\times 10^{21}$--$10^{22}$ atoms/cm$^{2}$  \\
$\mu$            & $0.5$                          \\
$\delta t$       & $50-250$~days                           \\
\hline
\end{tabular}
\end{center}
\label{input_parameters_A0620}
\end{table}
Figure~\ref{model0620-1} presents an example of the modeled light
curves, with $i=47^\mathrm{o}$, $d=0.66$~kpc, the bolometric luminosity 
$L_\mathrm{bol}(t=0)=0.25\, L_\mathrm{Edd}\,$, and 
$T_\mathrm{max}(t=35)=0.45$~keV (cf. Section~\ref{spectrum0620}). The
reduced $\chi^2$ for the X-ray light curve for $t\in[20,40]$~days is
1.17. 

We are not able to obtain a satisfactory fit to the slope of the optical
light curves. In principle, steeper optical light curves can be obtained
by taking into account irradiation of a thick or twisted disk. The outer
parts of the disk intercept some of the X-ray flux from the central
regions, causing the effective temperature of the outer disk, and
accordingly its flux, to increase. The intrinsic and reprocessed flux
depend on the accretion rate in different ways: the reradiated optical
flux decreases more rapidly, steepening the optical light curves.
\cite{esin_et2000} have modeled the outburst of \Mon{} taking into
account irradiation of the disk and assuming a relative half-thickness
for the disk of 0.12 (which is appreciably higher compared to the
standard model). In the standard $\alpha$--disk model, irradiation is
insignificant due to the small half-thickness of the disk, which is
typically about 0.03~\citep{lip-sha2000}. Probably, a contribution to
the optical flux from a disk, which is warped, should be taken into
account; a further study of the generation of optical radiation in a
time-dependent disk is necessary.

\Section{MODELING X-RAY NOVA \GUMus{} IN THE 1.2--37.2 keV 
X-RAY AND B AND V OPTICAL BANDS}

\subsection{Observations}

\subsubsection{X-ray Light Curves}

 The outburst of Nova Muscae (\GUMus, GU Mus) was discovered
independently by WATCH/GRANAT and ASM/GINGA (All-Sky X-Ray Monitor) on
January 9, 1991~\citep{lund_et1991, sunyaev1991}. The associated optical
outburst was also detected~\citep{lund_et1991}. For the modeling, we
use 1.2--37.2~keV data obtained with the GINGA Large Area
Counters~\citep{ebisawa_et1994}. The data in erg/(cm$^{2}$s) are
provided by the HEASARC database~\citep[see
Footnote~\ref{ftp-address}]{chen_et1997}. Following~\cite{chen_et1997},
we take the peak of the outburst to be on January 15, 1991, or JD
2448272.7862. 

The weighted least-squares regression line for data in the interval
$t\in [35,61]$~days yields $\ar=-0.0134\pm 0.0001$ and $\br=-6.511\pm
0.005$. However, the calculated reduced $\chi^2$ is very high, since the
observational errors are small and the data show an appreciable scatter
around the general trend of the model light curve. To select the models,
we use interval $(0.98-1.02)\,\ar$ for the the slopes of lines
tangent to the theoretical light curves at $t=48$~day.

\subsubsection{X-ray Spectrum at the Time of the Outburst}
As shown by~\cite{ebisawa_et1994},  after the peak the spectrum of 
\GUMus{} softened as the luminosity decreased. In~\cite{kitamo1992,
miyamo1993, ebisawa_et1994,   greine1994}, the observed X-ray spectrum was
approximated by a model with two components: a blackbody,
multi-color disk and a harder power-law component. Figure 15 in
\cite{ebisawa_et1994} indicates that, in the time interval of interest,
$T_\mathrm{in}$ for the spectral approximation \citep{ebisawa_et1994} is
roughly 0.7~keV.  

It was suggested by~\cite{kitamo1992} that 59\% of the flux on January
15 (near the outburst peak) was blackbody disk radiation, while the
remainder was contributed by a power-law component. During the following
25--30 days, the power-law component decayed more rapidly than the disk
component. ROSAT observations on January 25 (the 10th day after the
peak)~\citep{greine1994} suggested that the 0.3--20~keV flux at that
epoch was completely produced by the disk component. Using the
approximation for the observed 1.2--37.2~keV X-ray spectrum
from~\cite{miyamo1993} and the derived fluxes of the spectral
components, we conclude that the 1.2--37.2~keV flux during
$t\in[35,61]$~days after the outburst was determined  by the disk
radiation and can therefore be used in our modeling, since the
contribution from nondisk components is apparently negligible.

\paragraph*{Number of Hydrogen Atoms/cm$^2$ toward \GUMus{} and the
Color Excess} ~\newline 
In~\cite{cheng_et1992}, $\EBV\sim 0.29$ was
estimated from HST observations of the interstellar absorption 
profile at 2200\,\AA. \cite{shra-gonz1993} found $\EBV=0.3\pm0.05$ 
using a similar technique. In~\cite{dellav1991}, the same value
$\EBV=0.30\pm0.10$ was derived from interstellar Na D lines. Using
(\ref{hydrogen}), we arrive at $\NH \approx 1.4\times 10^{21}$
atoms/cm$^{2}$.

\cite{greine1994} obtained $\NH\approx 2.2\times 10^{21}$~atoms/cm$^{2}$
by modeling the combined ROSAT 0.3--4.2~keV and GINGA 1.2--37.2~keV data
for January 24--25 using a composite spectrum with a blackbody,
multi-color disk and a powerlaw component. For various multi-color disk
models, they obtained values from $1.7\times 10^{21}$ to $2.5\times
10^{21}$ atoms/cm$^{2}$.

\subsubsection{Optical Light Curves}
 We use the observational data from~\cite{king1996, dellav1998} and
derive weighted least-squares regression lines for the optical B and V
observations in the logarithmic flux units in the time interval 
$t\in[12,61]$~days. This yields $\ar=-0.0057\pm 0.0006$, $\br=-10.79\pm
0.03\,$ for the B band, and $\ar=-0.0052\pm 0.0006$, $\br=-10.98\pm
0.03\,$ for the V band. The corresponding reduced $\chi^2$ values are
2.8 and 7.3 for B and V, respectively, with 17 degrees of freedom. This
leads us to a conclusion similar to that for  \Mon{} optical
light curves (see Section~\ref{0620_optic}). We again assume that the
overall trend of the light curves can be described as quasi-exponential.

In model fitting, we use the observed color of the disk for
$t=48$~days: \hbox{B--V}$=0.27\pm 0.07^m$.

\subsection{Results for \GUMus}
 We compare  theoretical and observed curves in the interval $t\in[35,
61]$~days after the peak of the outburst. 
\begin{table}
\caption[Model parameters for\GUMus]{Input 
parameters for \GUMus{} models. The mass of the optical component
 $M_\mathrm{o}=0.8\, M_\odot$  is adopted from~\cite{chen_et1997}, 
and the binary period and
mass function  of the optical component $ 3.01\pm 0.15 M_\odot$ are 
from~\cite{orosz_et1996}.} 
\begin{center}
\begin{tabular}{c|l}
\hline
Parameter   &  Tested values \\
\hline
$M$              & $5-25\, M_\odot$                       \\
$M_{\mathrm{o}}$ & $0.8,0.9\, M_\odot$                    \\
$P$              & 0.433 days                              \\
$f(M_\mathrm{o})$& $ 3 \, M_\odot$                       \\
$\alpha$         & $0.1-1$                               \\
$\NH$ & $(1.4-2.5)\times10^{21}$ atoms/cm$^{2}$ \\
$\mu$            & $0.5$                               \\
$\delta t$       & $50-250$~days                           \\
\hline
\end{tabular}\\
\end{center}
\label{input_parameters_GUMus}

\end{table}
\begin{figure} 
  \begin{center}
  \resizebox{!}{0.6\textwidth}{\includegraphics[trim= 3cm 6cm 2cm 5cm, clip]{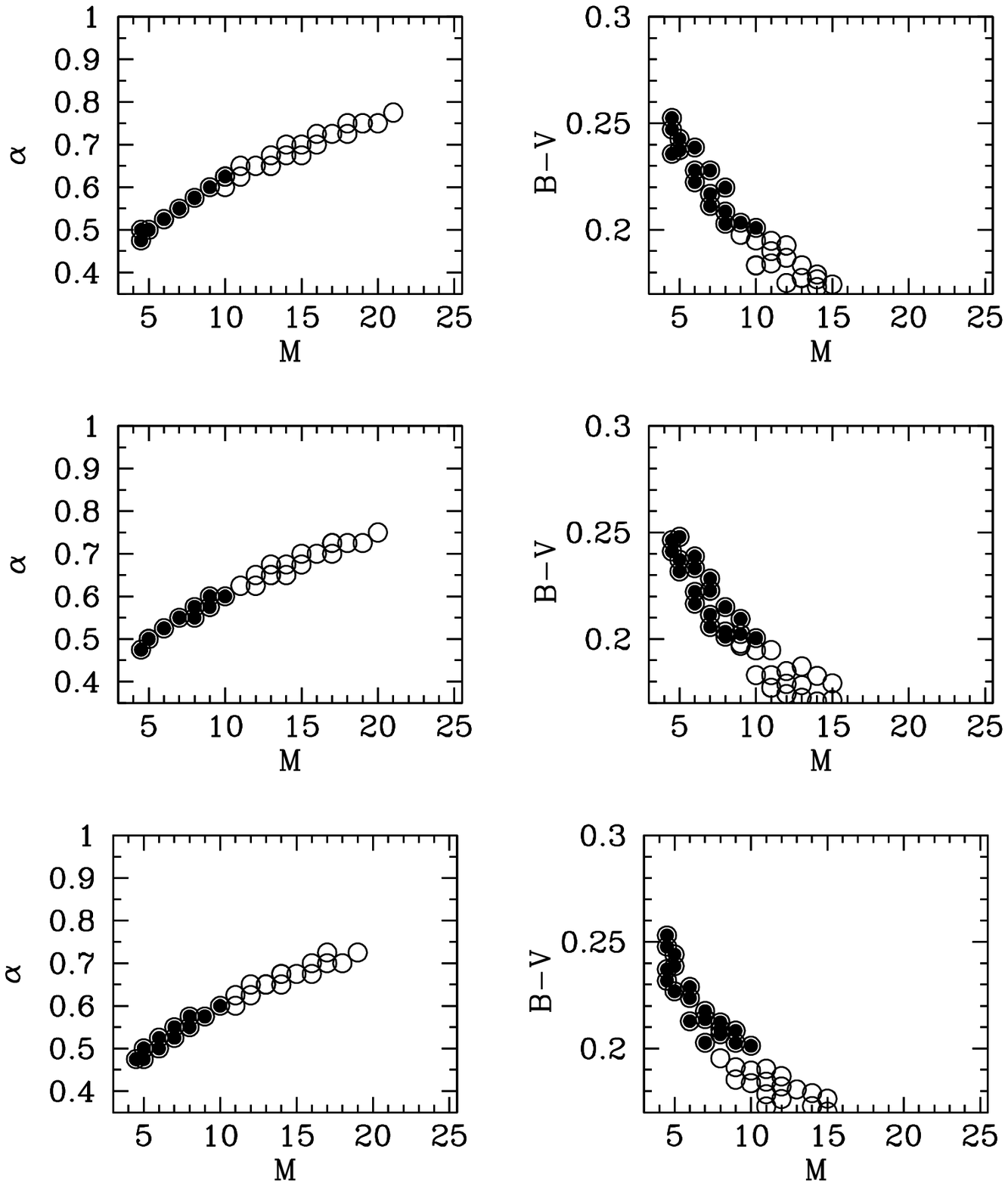}}
  \end{center}
  \caption{Modeled $\alpha$ and color of the disk of \GUMus{}
for the parameters in Table 4. The
mass of the optical component and the number of hydrogen 
atoms along the line of sight toward \GUMus{} vary from top to
bottom: $M_o = 0.8M_\odot$, $N_\mathrm{HI} = 1.4 \times
10^{21}$~atoms/cm$^2$ (two upper graphs), $M_o = 0.9 M_\odot$ , 
$N_\mathrm{HI} = 1.4 \times  10^{21}\mathrm{atoms/cm}^2$ (two
middle graphs), and $M_o = 0.8M_\odot$, $N_\mathrm{HI} = 2.5 \times
10^{21}\mathrm{atoms/cm}^2$ (two lower graphs). The filled circles 
satisfy the disk color condition $B-V = 0.27 ╠\pm 0.07$~mag for $t =
48$~days.
}
\label{modelsGUMus}
\end{figure}
Figure~\ref{modelsGUMus} presents the results of our modeling for the
parameters from Table~\ref{input_parameters_GUMus}. Comparing the lower
and upper left graphs, we can see a slight dependence of the 
results on the number of hydrogen atoms toward \GUMus. The resulting
$\alpha$ values for the \GUMus{} disk lie in the range 0.475--0.625 (for
small variations in $M_\mathrm{o}$ and $\NH$).

\begin{figure} 
  \begin{center}
   \resizebox{!}{0.35\textwidth}{\includegraphics{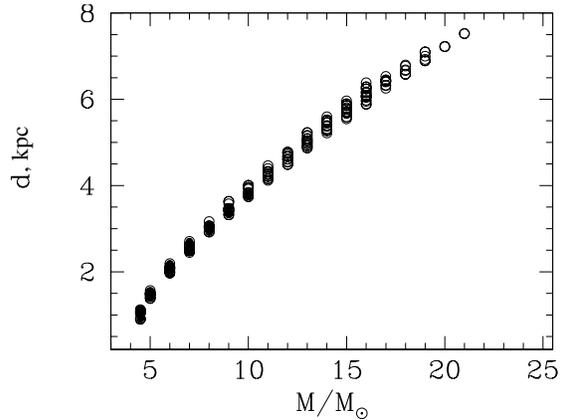}}
  \end{center}
  \caption{The distance-black hole mass relation for
\GUMus, modeled for the parameters from Table 4.
}
  \label{d-m1124}
\end{figure}
Figure~\ref{d-m1124} displays the dependence of the distance to 
\GUMus{} on the black hole mass. Estimates of the distance to  \GUMus{}
in the literature range from 1 to 8 kpc. For 3 kpc~\citep{cherep2000}, the
implied mass of the black hole in \GUMus{}  is $\sim 8\,M_\odot$
(Fig.~\ref{d-m1124}). The values for the black-hole mass obtained by us
are consistent with the observations (see~\cite{cherep2000}
and references therein). 

\begin{figure} 
  \begin{center}
   \resizebox{!}{0.35\textwidth}{\includegraphics{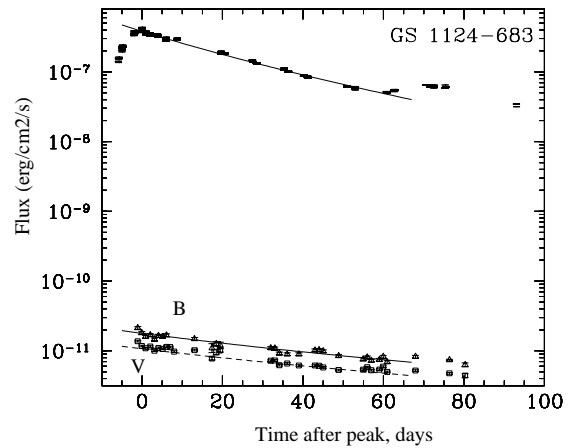}}
  \end{center}
  \caption{Example of modeling light curves of \GUMus{} in the 1.2--37.2~keV, 
B, and V band. The model parameters are
$M = 7M_\odot$, $M_o = 0.8M_\odot$, $\alpha=0.55$, $\delta t = 103$~days, 
and $N_\mathrm{HI} = 1.4 \times 10^{21} \mathrm{atoms/cm}^2$.}
  \label{modelGUMus-1}
\end{figure}

Figure~\ref{modelGUMus-1} presents model light curves for a particular
choice of parameters. In the corresponding model
$i=68^\mathrm{o}$, $d=2.6$~kpc, the bolometric luminosity
$L_\mathrm{bol}(t=0)=0.47\, L_\mathrm{Edd}\,$, and
$T_\mathrm{max}(t=48)=0.44$~keV. The model satisfactorily reproduces
both the X-ray and optical-light curves of \GUMus{}. 

\Section{CONCLUSION}

 We have modeled outbursts of two X-ray novae, \Mon{} and
\GUMus, using the model of a time-dependent $\alpha$--disk in a
binary system developed in~\cite{lip-sha2000}. 

The turbulence parameter $\alpha$ is estimated as 0.2--0.4 for \Mon{}
and 0.45--0.65 for \GUMus. The estimates of $\alpha$ are close to each
other, indicating the common nature for viscosity in the accretion
disks around compact objects, at least in these two sources. The value
of $\alpha$ ($\lesssim 1$) points out that the gas in the disks is
appreciably turbulent. We have also obtained relations between the
distances to the systems and the masses of the compact objects.

Thus, it is possible for the first time to model exponentially
decaying light curves of X-ray novae using a model with a thin accretion
disk with constant $\alpha$ and estimate value of $\alpha$.

Note that the optical B and V fluxes from \GUMus{} can be explained as
the disk radiation generated locally due to viscous heating. It is
likely that in the case of \Mon, we must also take into account the
contribution of reradiation by a warped disk. We will consider this
problem in a subsequent paper.

\Section{ACKNOWLEDGEMENTS}
 The authors are
grateful to V.F. Suleimanov for useful discussions. This work is
supported by the Russian Foundation for Basic Research (project nos.
01--02--06268, 00--02--17164), the ``Universities of Russia''
program (project no. 5559), and the State Science and Technology Project
``Astronomy'' (project 1.4.4.1). GVL is also grateful for financial
support from the ``Young Scientists of Russia'' program (www.rsci.ru,
2001). 
 Primary translation is done by K. Maslennikov.

\bibliographystyle{mn2e}

\end{document}